\documentclass[a4paper,10pt]{article}
\pdfoutput=1
\usepackage{mathptmx}
\usepackage{amssymb}
\usepackage{amstext}
\usepackage{ragged2e}
\usepackage{array}
\usepackage{color}
\usepackage{graphicx}
\usepackage{url}

\newcommand{\bq}{\begin{equation}}
\newcommand{\eq}{\end{equation}}

\newcommand{\GBS}{\mbox{GB/s}}

\newcommand{\GFS}{\mbox{GF/s}}

\newcommand{\LUPS}{\mbox{LUP/s}}
\newcommand{\MLUPS}{\mbox{MLUP/s}}
\newcommand{\GLUPS}{\mbox{GLUP/s}}
\newcommand{\GHZ}{\mbox{GHz}}

\newcommand{\WF}{\mbox{W/F}}
\newcommand{\bytes}{\mbox{bytes}}

\newcommand{\mus}{\mbox{$\mu$s}}
\newcommand{\eos}{~.}

\newcommand{\picfrac}{0.7}
\begin{document}
\title{Leveraging shared caches for parallel temporal blocking of stencil codes on multicore processors and clusters}
\author{Markus Wittmann, Georg Hager, Jan Treibig, Gerhard Wellein\\[2mm]
Erlangen Regional Computing Center\\University of Erlangen-Nuremberg\\
  Martensstr. 1, 91058 Erlangen, Germany}
\date{June 15, 2010}
\maketitle
\begin{abstract}
  Bandwidth-starved multicore chips have become ubiquitous.  It is
  well known that the performance of stencil codes can be improved by
  temporal blocking, lessening the pressure on the memory interface.
  We introduce a new pipelined approach that makes explicit use of
  shared caches in multicore environments and minimizes
  synchronization and boundary overhead. Benchmark results are
  presented for three current x86-based microprocessors,
  showing clearly that our optimization works best on designs with
  high-speed shared caches and low memory bandwidth per core.
  We furthermore demonstrate that simple bandwidth-based performance 
  models are inaccurate for this kind of algorithm and employ
  a more elaborate, synthetic modeling procedure.
  Finally we show that temporal blocking can be employed successfully 
  in a hybrid
  shared/distributed-memory environment, albeit with limited
  benefit at strong scaling.
\end{abstract}

\section{Introduction}

\subsection{The Jacobi stencil}

Stencil computations are central to many scientific and technical
applications, especially when solving partial differential equations
on regular lattices.  Even large-scale multigrid solvers employ
smoothing steps comprising stencil updates following the Gauss-Seidel
or Jacobi schemes~\cite{bergen05}\@.  Standard optimization techniques
like spatial blocking are usually applied to ensure optimal spatial
locality of data accesses. The advent of multicore chips
and advanced architectures like GPUs and the Cell processor
has recently initiated a new interest in stencil
optimizations~\cite{datta08,datta09}\@.

\begin{figure}[tbp]
\centering
\includegraphics*[width=\textwidth]{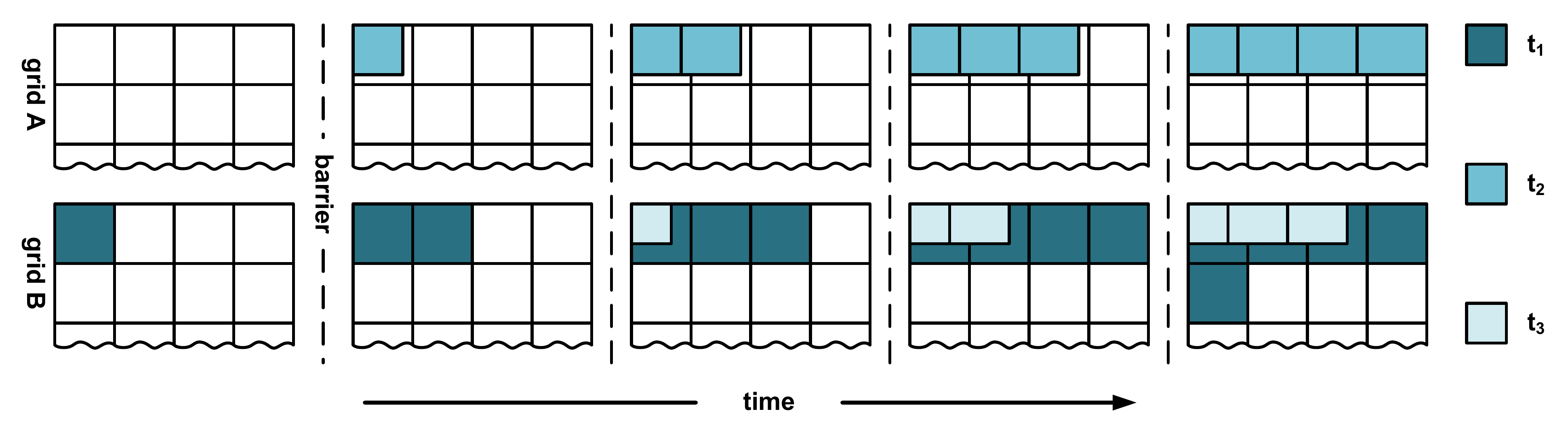}
\caption{\label{fig:pipelining}Temporal blocking by pipelining, shown
  here for three threads in two dimensions and separate grids $A$ and
  $B$\@.  All threads in a team update the most part of a block
  consecutively. Shifting the block by one cell in each direction
  after an update avoids extra boundary copies.  To avert race
  conditions, a global barrier is required after each block update.
  This scheme can be easily generalized to support multiple updates
  per thread. Possible optimizations include the use of a ``compressed
  grid'' update scheme, and relaxed synchronization (see text)\@.}
\end{figure}
The Jacobi algorithm is a simple method for solving boundary value problems.
Although it is not too useful
as a numerical algorithm by itself, it is the central component
of many multigrid solvers. Moreover it can serve as a prototype for
more advanced stencil-based methods like the lattice-Boltzmann
algorithm (LBM)\@. In three dimensions, one stencil (``cell'')
update is described by
\bq
 \label{eq:jacobi_kernel}
  B_{i,j,k}  =  \frac{1}{6}\left(A_{i-1,j,k} + A_{i+1,j,k} + A_{i,j-1,k}
             + A_{i,j+1,k} + A_{i,j,k-1} + A_{i,j,k+1}\right)\eos
\eq
Successive sweeps over the computational domain are performed to reach
convergence, writing to grids $A$ and $B$ in turn. Counting the
usually required cache line allocate on a write miss, the kernel
has a naive code balance of $B_\mathrm{c} = 8/6\,\WF$ (double
precision words per flop)\@. However, the actual number of words
transferred over the slowest data path (the memory bus) can be reduced
to three per stencil update with a suitable spatial blocking
scheme. Furthermore, the write allocate can
be circumvented on current x86-based
processors by employing \emph{non-temporal store} instructions,
which bypass the cache hierarchy.  These and other well-known standard
optimizations like data alignment and SIMD vectorization have been
applied for our baseline version of the algorithm and are described in
the literature \cite{datta08,datta09,wellein09}\@.
We use OpenMP-parallel C/C++  code whenever
possible, and revert to compiler intrinsics or assembly language
only if necessary, e.g., if the compiler refuses SIMD vectorization
because a loop
runs backward (as is the case with the compressed grid version
of pipelined temporal blocking, which will be described in
Sect.~\ref{sec:ptb})\@.

\subsection{Baseline and test bed}\label{sec:testbed}

The memory-bound performance of the baseline code on a given
architecture can be easily estimated by assuming that memory bandwidth
is the sole limiting factor, and that all other contributions can be
hidden behind it. This assumption is valid for current multicore
processors if all cores sharing a memory interface are used in a
parallel calculation, but may be false for single-threaded
code~\cite{thw09}\@. See Sect.~\ref{sec:perfmod} for a discussion
of this problem.

If the achievable STREAM COPY bandwidth (using non-temporal stores) is
$M_\mathrm{s}$, a ``perfect'' baseline Jacobi
code ($B_\mathrm{c}=0.33\,\WF$) should show a performance of
\bq\label{bwmodel}
P_0 = \frac{M_\mathrm{s}}{16\,\bytes}\,\left[\LUPS\right]\eos
\eq
We use the ``lattice site updates per second'' (\LUPS)
metric here.
Benchmark tests were performed on a variety of systems, which
are briefly described in Table~\ref{tab:architectures} together
with measurements for different low-level benchmarks, first and
foremost the STREAM COPY, \verb.A(:)=B(:). 
The remaining benchmark numbers were
obtained using a special array update loop, which is described
in Sect.~\ref{sec:perfmod}.

All systems are of ccNUMA type, so the usual precautions regarding
page placement apply. The Nehalem EX node is an early access (EA)
system provided by Intel, with half the memory boards removed.
Consequently, it has only half the main memory bandwidth of a fully
equipped system. It is worth noting that Nehalem EX features a
redesigned L3 cache with a ``banked'' structure, which leads to a very
high cache bandwidth. The consequences of this peculiarity will be
investigated in Sections~\ref{sec:perfmod} and \ref{sec:smresults}.

For benchmarking the distributed-memory algorithm described in
Sect.~\ref{sec:dmblock} we use a Nehalem EP cluster (node parameters
as above) with a fully nonblocking fat-tree QDR-InfiniBand network.


The idea behind temporal blocking is to perform multiple in-cache
updates on each grid cell before the result is evicted to memory,
thereby reducing effective code balance. Section~\ref{sec:ptb}
will introduce a pipelined temporal blocking
scheme in a shared-memory parallel context, while
Sect.~\ref{sec:dmblock} describes how and under what conditions these
optimizations can be put to use in a hybrid
(shared/distributed-memory) code.
\begin{table}[tbp]
  \caption{\label{tab:architectures}Overview on cache group structure,
    STREAM COPY (array size of 20,000,000 elements), and update performance
    for the systems in the
    test bed. Nontemporal stores were used for the STREAM COPY, so
    all bandwidth numbers denote actual bus traffic.}\par
\renewcommand{\arraystretch}{1.1}
\centering\begin{tabular}{rr>{\centering}m{1.5cm}>{\centering}m{1.5cm}>{\Centering}m{1.5cm}}
Name           &  &  \small\bfseries Nehalem EP    &   \small\bfseries Nehalem EX &  \small\bfseries Istanbul \\\hline
Type           &  &  Xeon X5550 @2.66\,\GHZ & Xeon X7560 @2.27\,\GHZ &   Opteron 2435 @2.6\,\GHZ \\\hline
Cores          &  &    4  &    8  &     6  \\
L1 size [kB]   &  &   32  &   32  &    64  \\
L2 size [kB]   &  &  256  &  256  &   512  \\
L3 size [MB]   &  &    8  &   24  &     5  \\\hline
L3 cache group [cores] &  &    4  &    8  &     6  \\
Sockets        &  &    2  &    4  &     2  \\\hline
$M_\mathrm s$   [\GBS]    && 19.0     &   7.9      &  10.5    \\
$M_\mathrm{um,1}$ [\GBS]  && 16.2     &   7.0      &   6.9    \\
$M_\mathrm{uc,1}$ [\GBS]  && 28.3     &  25.0      &  15.7    \\
$M_\mathrm{uc,max}$ [\GBS]  && 51.2     &  176.2      &  74.8 \\\hline
\end{tabular}
\end{table}

\subsection{Related work}

Improving the performance of stencil codes by temporal blocking is not
a new idea, and many publications have studied different methods in
depth~\cite{wonnacott00,jin01,kowarschik04,frigo05,kamil06}\@.
Conventional temporal blocking performs multiple updates on a small
block of the computational domain before proceeding to the next
block~\cite{kowarschik04}\@.
This strategy has the important drawback that in-cache stencil updates are not
naturally overlapped with loads and stores to main memory,
leading to under-utilization of the memory interface.
Cache-oblivious
algorithms as proposed by Frigo et al.~\cite{frigo05}
have the attractive property of being independent of
cache sizes, but they come at the cost of irregular block access patterns,
which cause many data TLB misses. This was shown for a 3D lattice
Boltzmann (LB) application kernel in Ref.~\cite{zeiser08}\@.

However, the
explicit use of shared caches provided by modern multicore CPUs has
not yet been investigated to great detail. Ref.~\cite{wellein09}
describes a ``wavefront'' method similar to the one introduced here.
However, that work was motivated mainly by the architectural peculiarities
of multi-core CPUs, and does not elaborate on specific optimizations
like avoiding boundary copies and optimizing thread synchronization.
Our investigation is more general and explores a much larger parameter
space. Finally there is, to our
knowledge, as yet no systematic analysis of the prospects of temporal
blocking on hybrid architectures, notably clusters of shared-memory
compute nodes.

\section{Shared-memory pipelined temporal blocking on shared caches}\label{sec:ptb}

\subsection{Algorithm and in-cache optimizations}

In contrast to previous approaches to cache reuse with stencil
algorithms, pipelined temporal
blocking makes explicit use of the cache topology on modern
processors, where certain cache levels are shared by groups of cores,
which we call \emph{cache groups}\@. On all test systems
used here (see Sect.~\ref{sec:testbed}), the outer cache level
(L3) is shared by all cores on a socket.

Pipelined blocking splits the set of all available threads into
\emph{teams} of size $t$, where a team runs on cores sharing a cache.
It is possible that the size of a team is smaller than the whole cache
group, but this option will not be explored here. Each team has one
``front'' thread, which performs the first $T$ updates on a certain
grid block (see Fig.~\ref{fig:pipelining} for a visualization with
three threads and $T=1$)\@. Of course the block is loaded to cache in
this process, and if it is small enough, the remaining threads can
perform further updates ($T$ each) on it in turn before it gets
evicted to memory. All threads in the team can be kept busy by keeping
this pipeline running, with different blocks in different stages until
each block of the whole computational domain has been updated $t\cdot
T$ times, which completes a \emph{team sweep}\@.
To avoid race conditions, the minimum distance between
neighboring threads is one block, but it may be larger. An estimate
for the maximum distance is given by the cache size divided by
$t$ times the size of one block. Due to the one-layer shift after each
block update, the actual amount of cache needed is actually larger,
depending on the blocksize and the overall number of updates,
$t\cdot T$\@. In the simplest case, the distance is kept constant
by imposing a global barrier across all threads after each block
update. See below for ways to reduce synchronization overhead.

Pipelined blocking can potentially overlap data transfer and calculation,
because the front thread continuously operates on new blocks, which
have to be fetched from memory. Compared to
the wavefront technique~\cite{wellein09}, it
does not incur extra work or boundary copies. In order to
achieve optimum in-cache performance, the kernel must be
fully SIMD-vectorized, i.e., packed arithmetic
(\verb.addpd./\verb.mulpd.) and packed loads and stores
are essential. Although the performance penalty for unaligned
moves from or to L1 cache is small on the most current x86 architectures,
we have also made sure that all loads and stores are aligned
to 16-byte address boundaries, and the corresponding instructions
are of the ``aligned'' type (\verb.movapd.).

The standard Jacobi algorithm iterates between two versions of the
computational grid, one holding the current state while the other is
used to store the next time step (i.e., the updated values). If the
temporally blocked algorithm performs an even number of time steps on
each block, the second grid becomes obsolete and can be reduced to a
collection of $t-1$ small temporary arrays, which must only be large
enough to hold all intermediate blocks handed down the
pipeline~\cite{wellein09}. Although this optimization reduces the
required data traffic from and to main memory by a factor of two,
it does not significantly cut down on in-cache transfers. However,
the well-known ``compressed grid''
optimization can be applied here: During the first team
sweep, each result is written to a location shifted by the vector
(-1,-1,-1) relative to its original position. In order to avoid
complex address calculations, alternate team sweeps shift by
(-1,-1,-1) and (+1,+1,+1), respectively, requiring reverse loops
(running from large to small indices) on all even sweeps.  Since the
compiler refuses to properly SIMD-vectorize the inner loop in this
case, SSE intrinsics were used to get optimal code.  The use of
non-temporal stores is unnecessary and even counterproductive; after
the $t\cdot T$ updates in a team sweep, a block gets evicted to memory
automatically by the usual replacement mechanisms. The benefit of
using ``compressed grid'' is that the second grid can be dropped
altogether (even including the small temporary arrays), saving nearly half the
memory. It also reduces in-cache traffic if at least two successive
layers of a block fit into the L1 cache.

In typical shared-memory systems there is usually more than one
outer-level cache group, the simplest case being a multi-socket node.
Hence, more than one team must be kept running.
We choose to use those additional threads to perform further updates
on the blocks already handled by the ``front'' team. This enlarges the
whole update pipeline to $n\cdot t\cdot T$ stages, where $n$ is the
number of teams. Since different teams do not share any cache, blocks
updated by one team must be transferred to another cache when the next
team takes over. As will be shown in the results section, it makes
sense to enforce a larger distance between successive teams than
between neighboring threads inside a team. We call this extra distance
the \emph{team delay}, $d_\mathrm{t}$\@.

A problem with our method of consecutive thread teams is that every
thread updates every block, rendering explicit (i.e., first touch)
ccNUMA placement optimizations mostly useless. However, since the
pressure on the memory interfaces is greatly reduced by the temporal
blocking, a round-robin page placement strategy is adequate to achieve
parallelism in memory access. Additionally, the distributed-memory
parallelization described in Sect.~\ref{sec:dmblock} can be used to
run one MPI process per socket, which alleviates the placement
problem.
The baseline Jacobi 
code does employ standard first-touch page placement.

\subsection{Relaxed synchronization}

\begin{figure}[tbp]
\centering
\includegraphics*[width=0.95\textwidth]{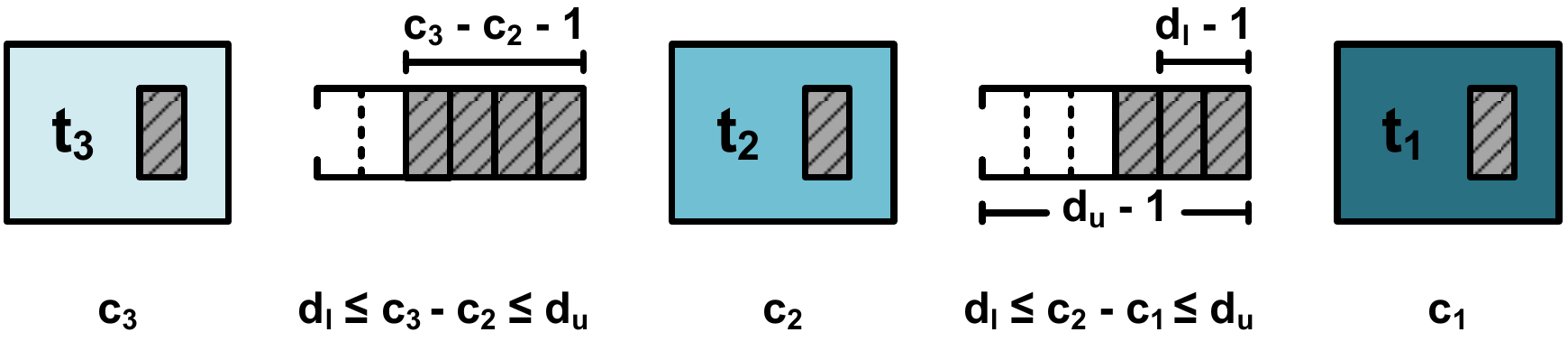}
\caption{\label{fig:relsync}Relaxed thread synchronization. A global
  barrier is avoided by observing ``soft'' lower and
  upper limits for the distance between consecutive threads.}
\end{figure}
It is well known that synchronizing a large group of threads running
on different cores can incur significant overhead. Depending on the
system topology (shared caches, sockets, ccNUMA locality domains) and on
the implementation used by the OpenMP compiler, a
barrier may cost hundreds if not thousands of cycles~\cite{thw09}\@.
With the number of cores per shared-memory node increasing steadily
over time, alternatives should be used where appropriate. In our
pipelined temporal blocking scheme, the barrier can be removed
completely if the minimum and maximum thread distance and team delays
are still observed. To this end, each thread $t_i$ maintains a
counter variable $c_i$, which carries the \verb.volatile. attribute
and is initialized to zero at the
start of each team sweep. It gets incremented whenever $t_i$ has
updated its current block. Each $c_i$ is located in a cache line of
its own to prevent false sharing. The conditions to allow thread
$t_i$ to start updating the next block are then
\bq\label{eq:updcond}
c_{i-1}-c_i\geq d_\mathrm{l}\quad\wedge\quad c_i-c_{i+1}\leq d_\mathrm{u}\eos
\eq
The first condition averts data races, whereas the second maintains a
maximum distance between consecutive threads. The team delay is
trivially implemented by adding $d_\mathrm{t}$ to $d_\mathrm{l}$ on a
team's front thread and to $d_\mathrm{u}$ on its ``rear''
thread. Overall front and rear threads (i.e., the front thread of the
first and the rear thread of the last team) ignore the first and the
second condition, respectively.

In this scheme, only thread $t_i$ updates its own counter $c_i$; all
others read its updated value by means of the standard cache coherence
mechanisms.  Making the variables volatile prevents register
optimizations.
The naive choice of
$d_\mathrm{l}=d_\mathrm{u}=1$ imposes a rigid lock-step between threads.
As will be shown in the results section, it is better to choose a different
value at least for $d_\mathrm{u}$, allowing for some
``looseness'' in the pipeline.

In pseudo-code the complete
algorithm for one team sweep, including the compressed grid and
relaxed synchronization
optimizations, looks as follows (for a team delay of zero):
\begin{verbatim}
/*  1 */  int direction = 1; // or -1, if backward sweep
/*    */  int shiftVector[] = {-1,-1,-1};
/*  3 */  
/*    */  #pragma omp parallel
/*  5 */  {
/*    */    int blockCounter = 0;
/*  7 */    int tIdx = omp_get_thread_num();
/*    */  
/*  9 */    while(block = blocks.NextBlock(direction,blockCounter))
/*    */    {
/* 11 */      ++blockCounter;
/*    */      while(c[tIdx]+d_l>=c[tIdx-1]); // spin
/* 13 */      focusVector[] = {tIdx*T, tIdx*T, tIdx*T};
/*    */  
/* 15 */      for (i = 0; i < T; ++i) {
/*    */        update_block(
/* 17 */          source = block.Coords()+direction*focusVector[],
/*    */          dest = block.Coords()+direction*   // compressed
/* 19 */               (focusVector[]+shiftVector[]) // grid
/*    */        );
/* 21 */        focusVector[]+=direction*shiftVector[];
/*    */      }
/* 23 */  
/*    */      if (blockCounter == totalBlocks)
/* 25 */        c[tIdx] += d_u + 1; // pipeline wind-down
/*    */      else {
/* 27 */        ++c[tIdx];
/*    */        while(c[tIdx]>c[tIdx+1]+d_u); // spin
/* 29 */      }
/*    */    }
/* 31 */  } // end parallel
\end{verbatim}
The \verb.update_block(). function (lines~16--20)
performs the actual Jacobi
update on a block; the coordinates to use for reading and
writing are given in the \verb.source. and \verb.dest.
parameters, respectively. Lines~12 and 28
contain the spin-waiting loops that implement the correctness
and cache size conditions (\ref{eq:updcond}), respectively.
\verb.focusVector[]. stores the coordinates of the current
block to be updated, and is updated in line~21 to reflect
the shift shown in Fig.~\ref{fig:pipelining}\@.
For a backward sweep, \verb.direction.
is set to \verb.-1.; everything else stays the same.

\subsection{Performance predictions}\label{sec:perfmod}

In order to get an estimate on the expected performance
gain from pipelined temporal blocking, it is essential to identify
the bottlenecks that limit scalability on a shared cache.
A viable assumption would be that a modified version of
(\ref{bwmodel}) is valid even if $t$ updates are performed
in cache per stencil (we set $T=1$):
\bq\label{bwmodel2}
P_0(t) = \frac{t\cdot M_\mathrm{um,1}}{16\,\bytes}\,\left[\LUPS\right]\eos
\eq
Here, $M_\mathrm{um,1}$ is the single-thread bandwidth for a 
single-stream update benchmark:
\begin{verbatim}
  for(int i=0; i<N; ++i) // N large
    a[i] += 1.0;
\end{verbatim}
This loop models  the data transfer behavior of a compressed
grid (and probably temporally blocked) Jacobi solver if we can assume
that at least two successive layers of a block fit into a cache level
further up in the hierarchy. Then, $t$ updates cause 16 bytes of 
memory data traffic per stencil.
\begin{figure}[tbp]
\centering
\includegraphics*[width=\picfrac\columnwidth]{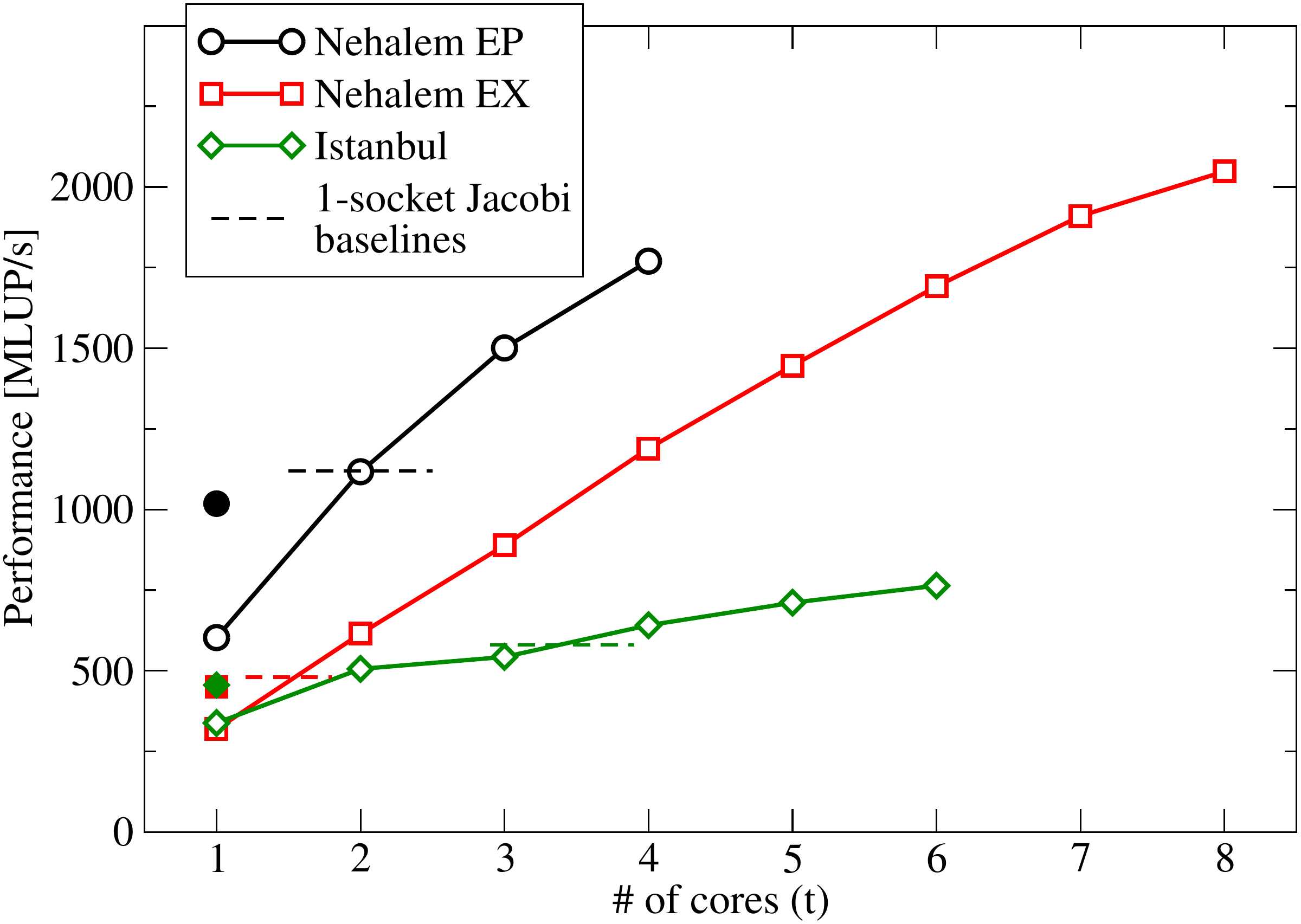}
\caption{\label{fig:ppscaling}Performance and scalability of the pipelined
   Jacobi solver, using relaxed synchronization with $T=1$ within an
   L3 cache group on all three benchmark architectures. Dashed lines indicate
   the performance of the best standard Jacobi implementation using all
   threads on a socket. Filled symbols denote the $t=1$, $T=1$ prediction
   from a single-threaded update benchmark.}
\end{figure}
Figure~\ref{fig:ppscaling} shows performance and scalability
for the $T=1$ case within an L3 cache group. The one-socket baseline
for the ``best'' standard Jacobi code is indicated and shows how
many cores are required at $T=1$ to match its performance with
pipelined temporal blocking. Filled symbols show the prediction
based on the bandwidth model (\ref{bwmodel2}), using the single-threaded
update bandwidth ($M_\mathrm{um,1}$ in Table~\ref{tab:architectures}). 
The measurements
clearly fall short of this prediction, especially on the Nehalem EP.
Our assumption that raw memory bandwidth is always the limiting factor 
must thus be false. A more careful analysis is in order that takes into
account the actual in-cache runtime of the kernel. We shall see
that, contrary to popular wisdom, the cache hierarchy and the
code execution from L1 cache are 
not infinitely fast even though the code balance of a particular
kernel seems to support this assumption.

\begin{figure}[tbp]
\hspace*{\fill}\includegraphics*[width=0.4\columnwidth]{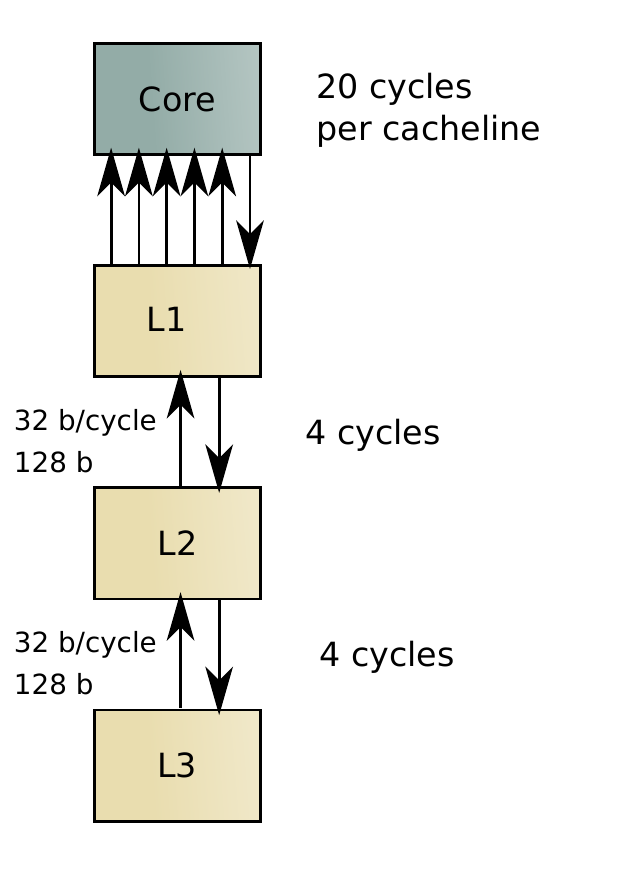}\hspace*{\fill}%
\includegraphics*[width=0.4\columnwidth]{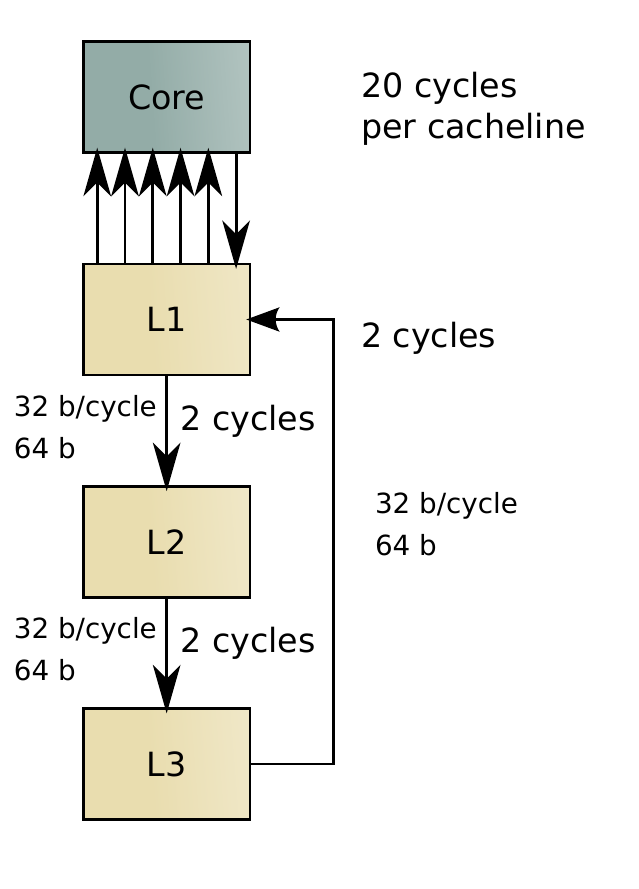}\hspace*{\fill}
\caption{\label{fig:CLmodel} In-cache performance analysis of the
compressed grid
Jacobi kernel on Intel Nehalem EP (left) and AMD Istanbul (right).
Each arrow represents one 64-byte cacheline transfer. The number of cycles required
for all transfers between each pair of cache levels is indicated on the 
right. Those contributions may partially overlap; see text for details.}
\end{figure}
The given seven-point Jacobi stencil performs six flops (one multiplication and five
additions), six loads (five loads if assuming that variables on the central line can
be kept in register) and one store. The Intel Nehalem EP can execute one SIMD
multiply, one SIMD add, one 16-byte load and one 16-byte store
instruction per cycle. Under the assumption of optimal scheduling and
pipelining this results in a minimal runtime of five cycles per vector update (20
cycles for a cacheline update) on data in the L1 cache. The
performance is limited by the arithmetic instruction throughput as well as by
the load/store capabilities. The ``lightspeed'' performance on our benchmark
system is thus 6.4\,\GFS\ (1.1\,\GLUPS),  so the
absolute limit for any kind of optimized Jacobi solver is 4.4\,\GLUPS\ per
socket, or a speedup of four compared to the standard Jacobi baseline
(see Fig.~\ref{fig:ppscaling}). If the data set does not fit in
the L1 cache, additional cacheline transfers are necessary. On Intel processors
the caches are inclusive, which means that a valid cacheline is duplicated in all
``lower'' cache levels. Figure~\ref{fig:CLmodel} (left) illustrates how the 
necessary data
transfers inside the cache hierarchy contribute to the total runtime if
one can assume that at least two successive layers of a block fit into
the L1 cache, and blocks are handed down the pipeline through L3. For a
detailed introduction to this kind of intra-cache model cf.~\cite{thw09}.  
The estimated runtime for the L3 cache domain is 28 cycles per cacheline update.
Measurements show that the Intel Nehalem processor is able to partially overlap the
reloads from L3 to L2 cache with the cycles spent on executing the instructions
on data in the L1 cache (however, refills from L2 to L1 can \emph{not} overlap
with L1-register traffic).  Hence, the prediction for the duration of 
one cacheline update (eight successive stencils) in L3 is 24--28 cycles, or,
in terms of L3 bandwidth, $B_\mathrm j={}$12.2--14.2\,\GBS. Note that the array update benchmark
described above runs in 4.7 cycles in L1 (with a theoretical
minimum of 4.0) and 12 cycles in L3 per cacheline update. The latter
number can be derived from the single-threaded in-L3 bandwidth of
the update benchmark shown in Table~\ref{tab:architectures} 
($M_\mathrm{uc,1}$). The Nehalem EP and EX CPUs show very similar in-cache
behavior for single-threaded execution.

The same scheme can be applied for the AMD Istanbul processor, where
our Jacobi stencil is limited by arithmetic instruction throughput
(also one SIMD multiply and one SIMD add per cycle). In contrast
to the Intel processors, AMD uses an exclusive cache design: L2 
and L3 are victim caches and all data loads go directly 
to the L1 cache.  The resulting transfer
paths and volumes are illustrated in Fig.~\ref{fig:CLmodel}.  The estimated
minimum runtime for the compressed grid Jacobi stencil in L3 is 26 cycles, which is
equivalent to a bandwidth of $B_\mathrm j={}$12.8\,\GBS\ on our test machine. Our
array update benchmark requires only 18 cycles in L3.
Note that the cycle predictions of the in-cache model are lower limits,
and even more time is needed per cacheline update in practice. However,
they yield a much better prediction than pure bandwidth considerations
(L2 and L3 caches often have the same \emph{nominal} bandwidths as
L1). 

Hence, the conclusion for the single-threaded case is that the in-cache
operations are a major contribution to overall runtime, and cannot
be ignored. Since each L1 and L2 cache is assigned to a single core on all
architectures in the test bed, the next bottleneck to consider is the
shared L3 cache. We assume that it can deliver a maximum aggregate 
bandwidth to all agents (main memory and cores), and becomes a performance
constriction if the sum of all bandwidths hits the maximum. We consider 
$M_\mathrm{uc,max}$, the aggregate bandwidth for the update benchmark
using all cores in the L3 group, to be the upper limit for L3.
See Table~\ref{tab:architectures} for performance data.

If $t$ threads are used for pipelined temporal blocking via L3, all of
them exert pressure on this cache according to their execution timings
as described above. The required memory transfers for bringing a block
into cache and for its final eviction add another read and write stream,
so that the required bandwidth to reach full scalability is
$(t+1)B_\mathrm j$. For Nehalem EP, if we assume $B_\mathrm j\approx 10\,\GBS$,
we expect from the relatively small ratio of $M_\mathrm{uc,max}/B_\mathrm j$ 
that pipelined blocking will only just scale up to $t=4$ threads.
On the other hand, Nehalem EX should scale well up to eight threads
and beyond. Indeed, we see parallel efficiencies around 80\% on both
chips up to $t=4$ and $t=8$, respectively (see Fig.~\ref{fig:ppscaling}). 
In contrast, the AMD Istanbul processor shows mediocre scalability but
no typical saturation behavior. As will be shown in the next section,
$T=1$ is not an optimal choice here; more updates per block are needed
to reach adequate performance. The reason for this is unclear and
also unexpected from the low-level performance data we have collected.


\subsection{Socket and node results}\label{sec:smresults}

\begin{figure}[tbp]
\centering
\includegraphics*[width=0.9\columnwidth]{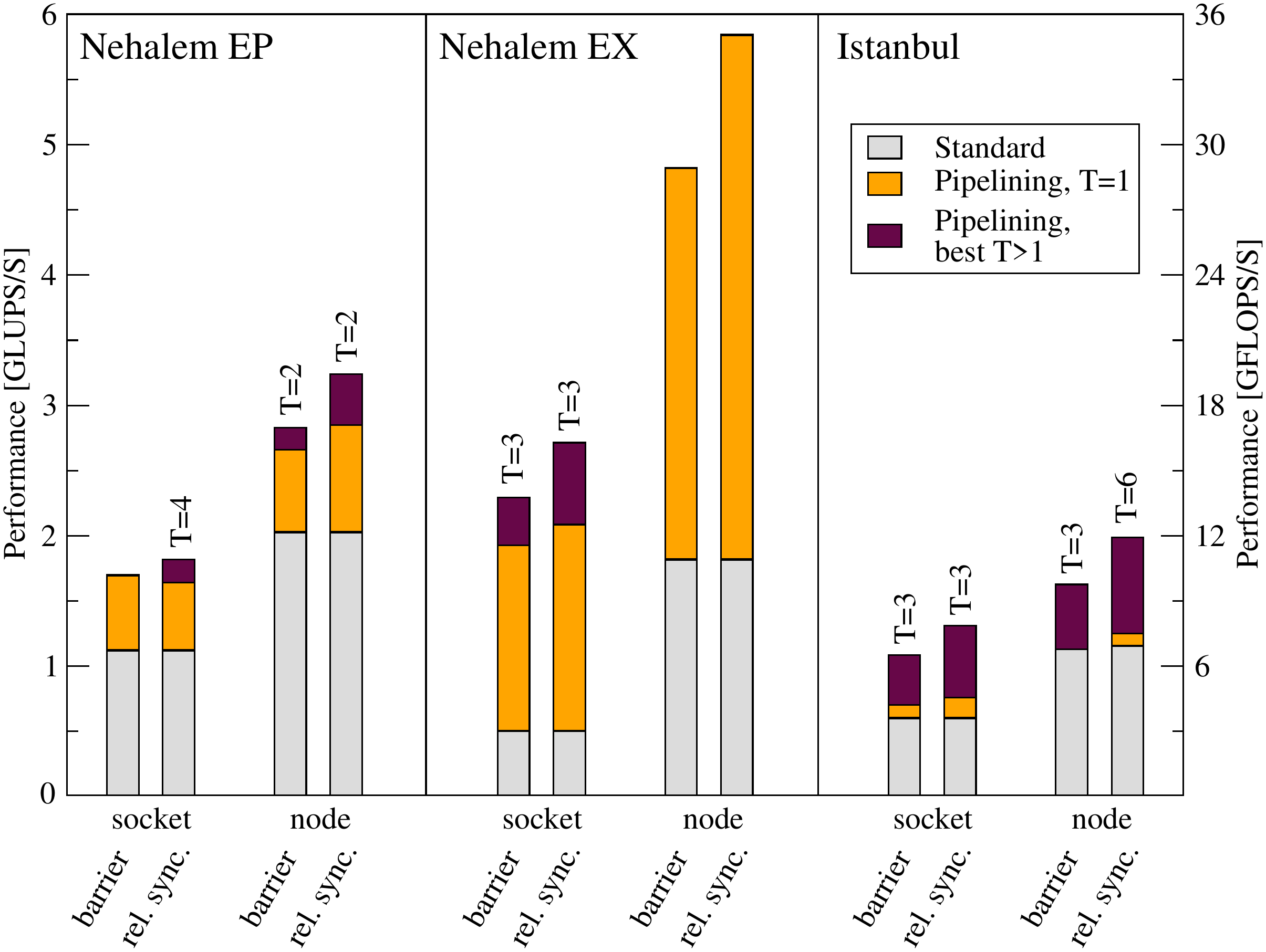}
\caption{\label{fig:node}
      	Single-socket and single-node performance results for
      	different variants of pipelined temporal blocking versus the
      	standard version ($600^3$ grid)\@. $T=1$ and best $T>1$
      	performance, and the gain from relaxed synchronization are
	indicated. Optimal values for $T$ were
      	determined empirically. }
\end{figure}
We must stress that the parameter space for temporal blocking schemes,
and especially for pipelined blocking, is huge. The optimal
choices reported here have been obtained experimentally by
parameter sweeps, with some guidance from experience with older codes.

Figure~\ref{fig:node} shows single-socket (one team)
and node results (two teams) on our test systems for a fixed problem
size of $600^3$ and different algorithmic variants:
the standard Jacobi solver baseline, pipelined temporal blocking
with $T=1$ and $t$ being equal to the number of cores in a socket
(or system), and pipelined temporal blocking with $T$ chosen
optimally. For all versions, the difference between
a global barrier and our relaxed synchronization scheme is also
indicated.
The standard Jacobi data was  obtained with a blocksize
of roughly $600\times 20\times 20$ ($b_x\times b_y\times b_z$); it is
well known that due to the hardware prefetching mechanisms on current
x86 designs, a long inner loop (comparable to the page size) is
favorable~\cite{datta08}\@.  While the results are rather insensitive
to the blocksizes in $y$ and $z$ directions as long as the cache
size restrictions are observed, the inner loop length is also
decisive for good performance on the temporally blocked versions,
and depends on $T$\@.
Best performance is achieved around $b_x\approx 120$ if
$T>1$, in contrast to the $T=1$ case where $b_x$ should be chosen
as large as possible.


The possible gain from pipelined temporal blocking varies strongly
across the benchmark architectures. As expected, a large speedup is
achieved if the gap between a socket's memory bandwidth and its L3
saturation bandwidth is large. Nehalem EP is exceptional in the sense
that this discrepancy is relatively small; also, using more cores
with the same L3 design (as implemented in the Intel ``Westmere'' chip)
would not make sense at all.  This makes temporal blocking
schemes of any kind less beneficial than on other, more
bandwidth-starved architectures.  With a speedup of 5--6 on the
socket and of 3 on the full node, Nehalem EX clearly dominates due to
its scalable L3 cache design, large number of cores, and (in our EA
system) low memory bandwidth. Both Intel processors get most of the
speedup already at $T=1$, whereas the AMD Istanbul  requires
more time steps per block. A possible reason could be inefficient
hardware prefetching within the cache hierarchy~\cite{thw09};
latencies have less impact if more work is done before fetching
the next block. 

The benefit of relaxed synchronization naturally goes up with the
number of cores and sockets in the team, and gets as large as 15\% on
Nehalem EX. Hence, we expect efficient synchronization schemes to
become more and more important in the future as the core count per socket
increases. It is certainly possible to achieve even larger gains by
choosing a smaller (and then non-optimal) block size.



\begin{figure}[tbp]
\centering
\includegraphics*[width=0.6\columnwidth]{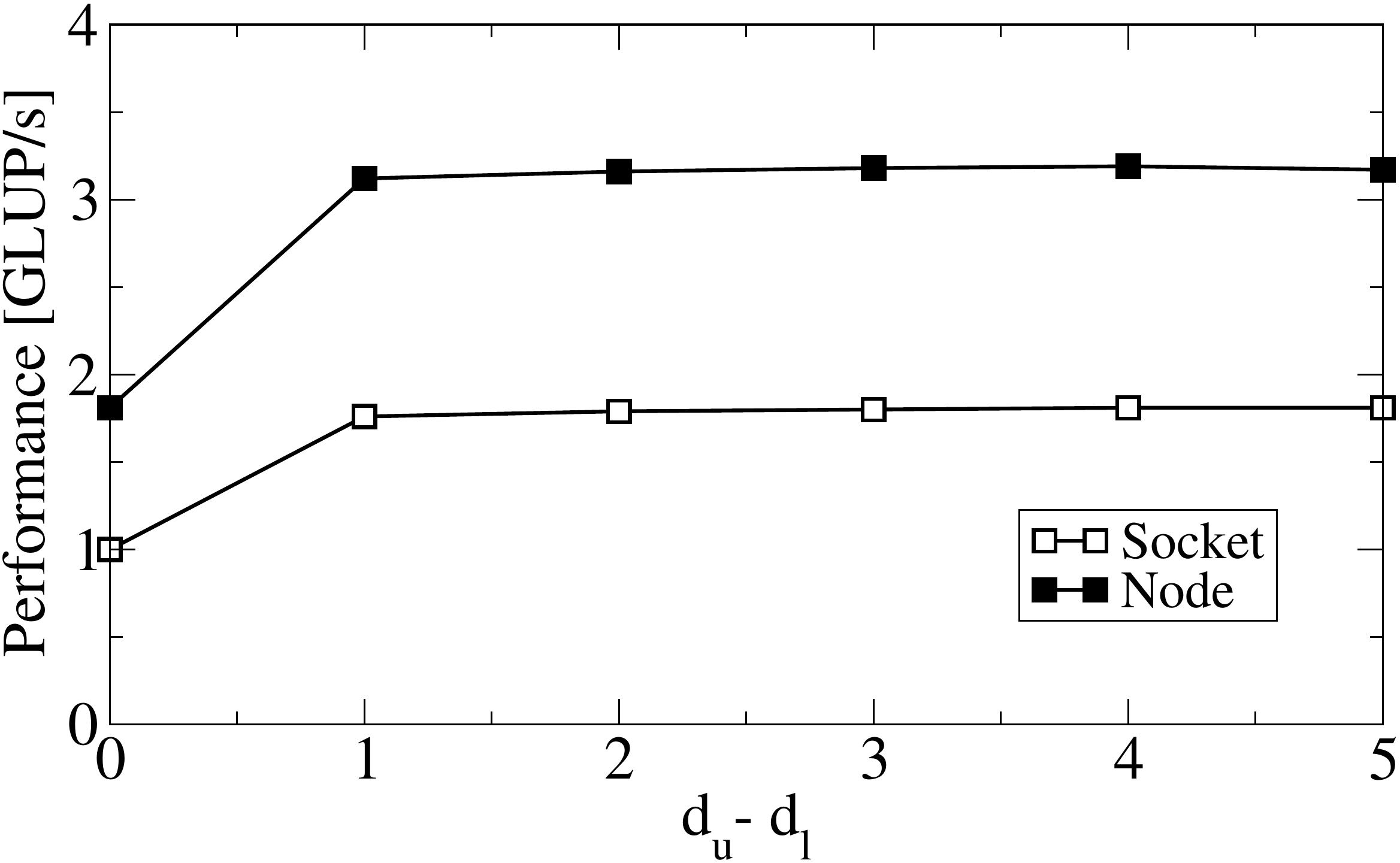}
\caption{\label{fig:loose}Influence of pipeline looseness on the socket
  (open symbols) and node level (filled symbols)\@. The case
  $d_\mathrm{u}-d_\mathrm{l}=0$ represents a rigid ``lockstep'',
  which is obviously hazardous. Data was taken on Nehalem EP, 
  but the general characteristic is very similar on all architectures.}
\end{figure}
The optimal range of values for $d_\mathrm{u}$, the upper limit for the
distance of neighboring threads, was determined to be 2--4 with the
block sizes chosen, largely 
independent of architecture (see Fig.~\ref{fig:loose})\@.
This allows for sufficient looseness in the
pipeline without running the danger of blocks falling out of cache
before the team's rear thread has done its updates on them. Compared
to the ``lockstep'' case $d_\mathrm{l}=d_\mathrm{u}=1$, a performance
gain of about 80\% can be observed for Nehalem EP. Of course,
$d_\mathrm{u}$ and the blocksize are strongly coupled, and larger
blocks would require smaller $d_\mathrm{u}$, but we could not find
better combinations than the ones reported here.
A finite team delay $d_\mathrm{t}$
only has a very slight impact on this architecture (about 3\%
improvement for $d_\mathrm{t}=8$), and its influence will not
be studied further.

Scalability of the pipelined code is not perfect across sockets, since
proper NUMA placement cannot be enforced, as described above. However,
distributed-memory parallelization (see next section) can employ one
process per socket, eliminating the need for first-touch parallel
placement.

\section{Distributed-memory parallelization}\label{sec:dmblock}

\subsection{Multi-halo exchange}

\begin{figure}[tbp]
\centering
\includegraphics*[width=0.45\columnwidth]{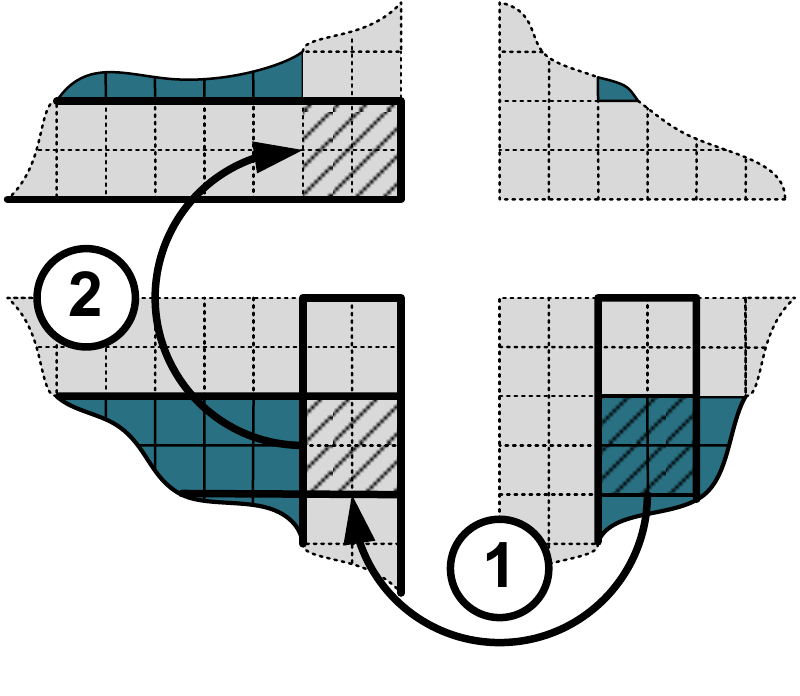}
\caption{\label{fig:dce}Multi-layer halo communication. Each
  halo is transmitted consecutively along the three coordinate
  directions, avoiding direct communication across edges and
  corners~\cite{ding01}.}
\end{figure}
The parallel temporal blocking schemes described above work on
multicore-based shared-memory architectures. Stencil algorithms are
usually straightforward to parallelize on dis\-tri\-bu\-ted-memo\-ry
systems using domain decomposition and halo layer exchange, and the
temporally blocked Jacobi code is no exception: The computational
domain is decomposed as usual, but instead of a single halo, $h$
layers must be exchanged after $h=n\cdot t\cdot T$ updates have been
performed per subdomain.  
Subdomains overlap by $h-1$ grid layers, and extra work is
involved on the boundaries because update number $s$ covers a domain
that is $h-s$ layers larger in each direction. The amount of data
communication per stencil update is roughly the same as for
no temporal blocking, except for edge and corner contributions,
which only become important on very small subdomains (see below)\@.
An important side effect of multi-layer halo exchange is that
communication takes place across subdomain edges and corners.
Latency-dominated small messages can be avoided by transmitting halos
consecutively along the three coordinate directions~\cite{ding01} (see
Fig.~\ref{fig:dce})\@.
\begin{figure}[tbp]
\centering
\includegraphics*[width=\picfrac\columnwidth]{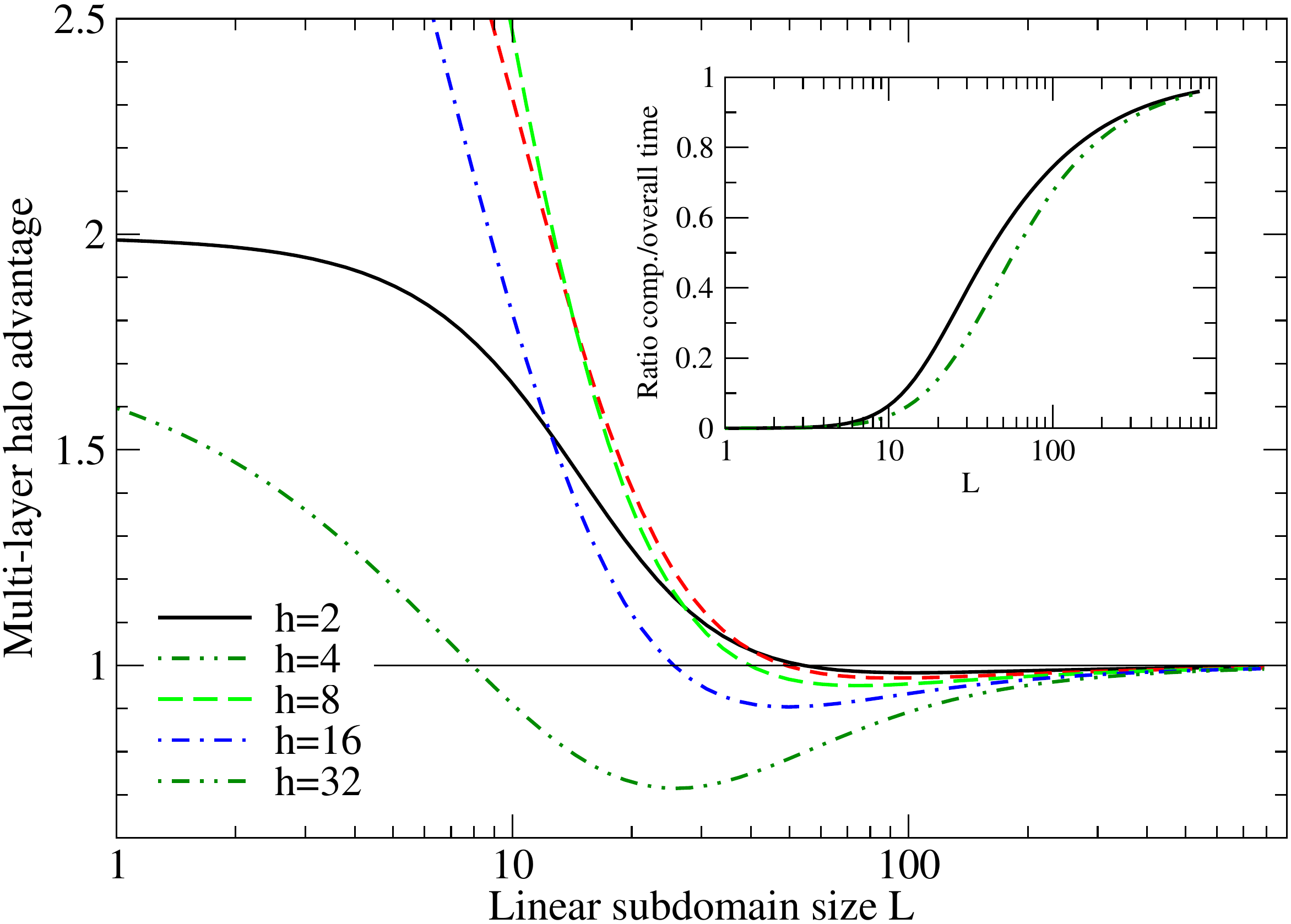}
\caption{\label{fig:mlh-adv}Theo\-re\-ti\-cal multi-layer halo advantage
  versus linear subdomain size $L$ for different halo widths
  $h$\@. Parameters are set for a vector-mode hybrid Jacobi solver on
  a QDR-IB network and a per-node performance of 2000\,\MLUPS\ (see
  text for details)\@. Inset: Ratio of computation versus overall time
  (``computational efficiency'') for the corner cases $h=2$ and
  $h=32$\@.}
\end{figure}

The question arises whether the use of multi-layer halo exchange has
any significant impact on performance. Two factors lead to opposite
effects here: If the subdomain surface area is very small, aggregating
multiple messages into one may be beneficial; effective bandwidth
rises dramatically with growing message size in the latency-dominated
regime. On the other hand, the surface to volume ratio is large in
this limit, leading to a significant overhead from communication and
extra halo work.
The different contributions to execution time (``bulk'' and additional
``face'' stencil updates, and halo exchange) can be calculated,
assuming a simple latency/bandwidth model for network communication
and no overlap between calculation and data transfer.

While only simple algebra is involved, the resulting expressions are
very complex, so we restrict ourselves to a graphical analysis. The
main panel of Fig.~\ref{fig:mlh-adv} shows the predicted ratio of
execution times between a standard one-layer halo version and
$h$-layer exchange for cubic subdomains of size $L^3$ and different
$h$\@. We have set the parameters for a QDR-InfiniBand network here,
with an asymptotic (large-message) unidirectional bandwidth of 3.2\,\GBS\ and a
latency of 1.8\,\mus\@. The single-node performance was assumed to be
2000\,\MLUPS, independent of $L$ (which only roughly holds in
practice)\@. As expected, multi-layer halos have no influence at large
subdomain sizes.  As the domain gets smaller ($20\lesssim L\lesssim
100$), extra halo work starts to degrade performance, but a relevant
impact can only be expected at $h\gtrsim 16$\@. At even smaller
$L\lesssim 20$, the positive effect of message aggregation
over-compensates the halo overhead, leading to substantial performance
gains. Although this looks like a good result, the ratio of
computation time versus overall execution time as shown in the inset
of Fig.~\ref{fig:mlh-adv} proves that the algorithm is strongly
communication-limited below $L\approx 100$, such that parallel
efficiency is very low. Any gain obtained by sophisticated
temporal blocking is squandered by communication overhead in this
limit.

Note that this simple model disregards some important effects like
switching of message protocols, overhead for copying to and from
message buffers, load imbalance, etc. Its purpose is to get a rough
idea about where to expect benefits from distributed-memory
parallelization of pipelined blocking with multi-layer halos.

\subsection{Distributed-memory results}\label{sec:dmresults}

\begin{figure}[tbp]
\centering
\includegraphics*[width=\picfrac\columnwidth]{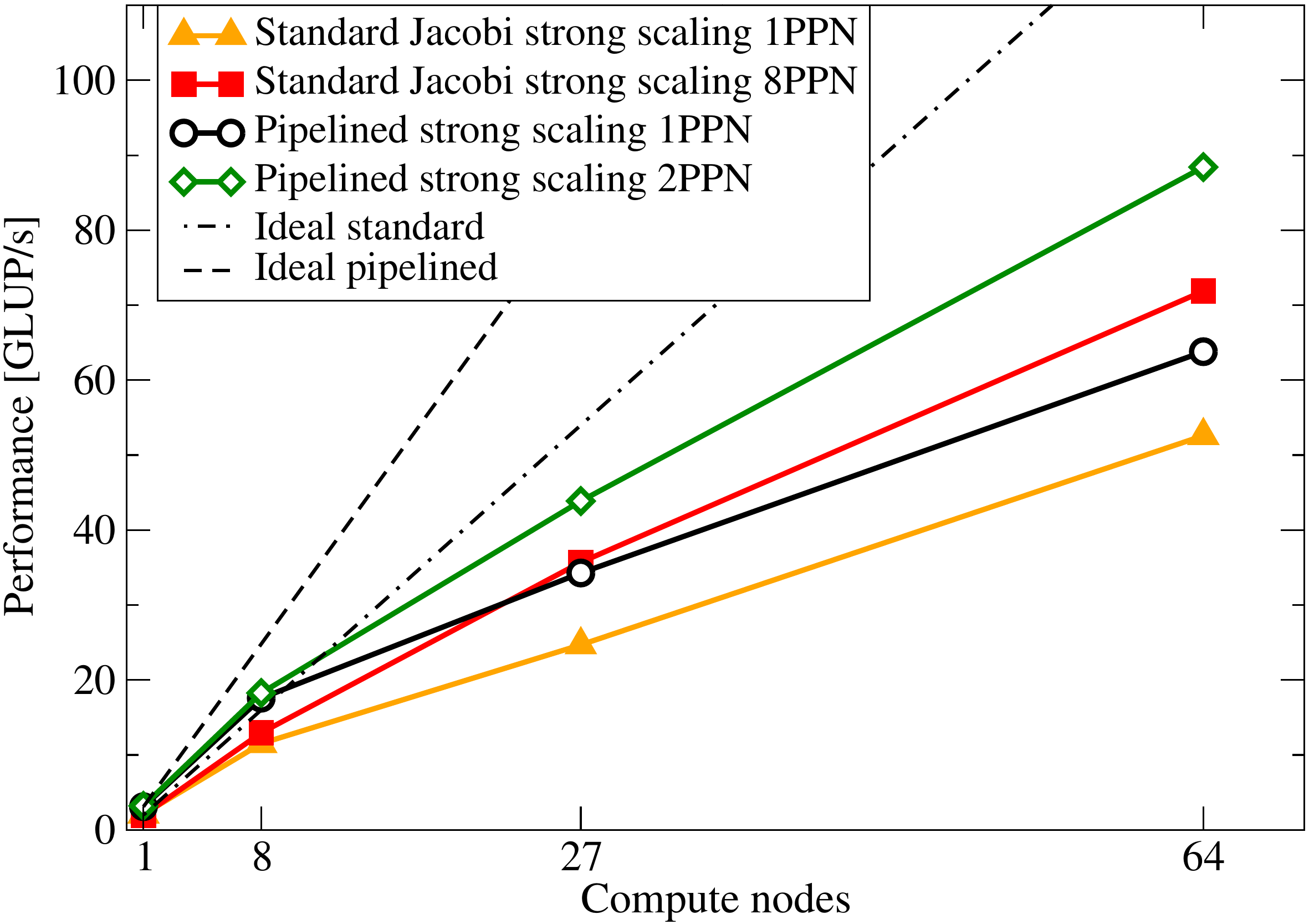}
\caption{\label{fig:dm-strong}Distributed-memory parallel performance
	(strong scaling) of the standard and the multi-halo
        pipelined Jacobi solvers with relaxed
	synchronization, at a problem size of $600^3$\@.}
\end{figure}
\begin{figure}[tbp]
\centering
\includegraphics*[width=\picfrac\columnwidth]{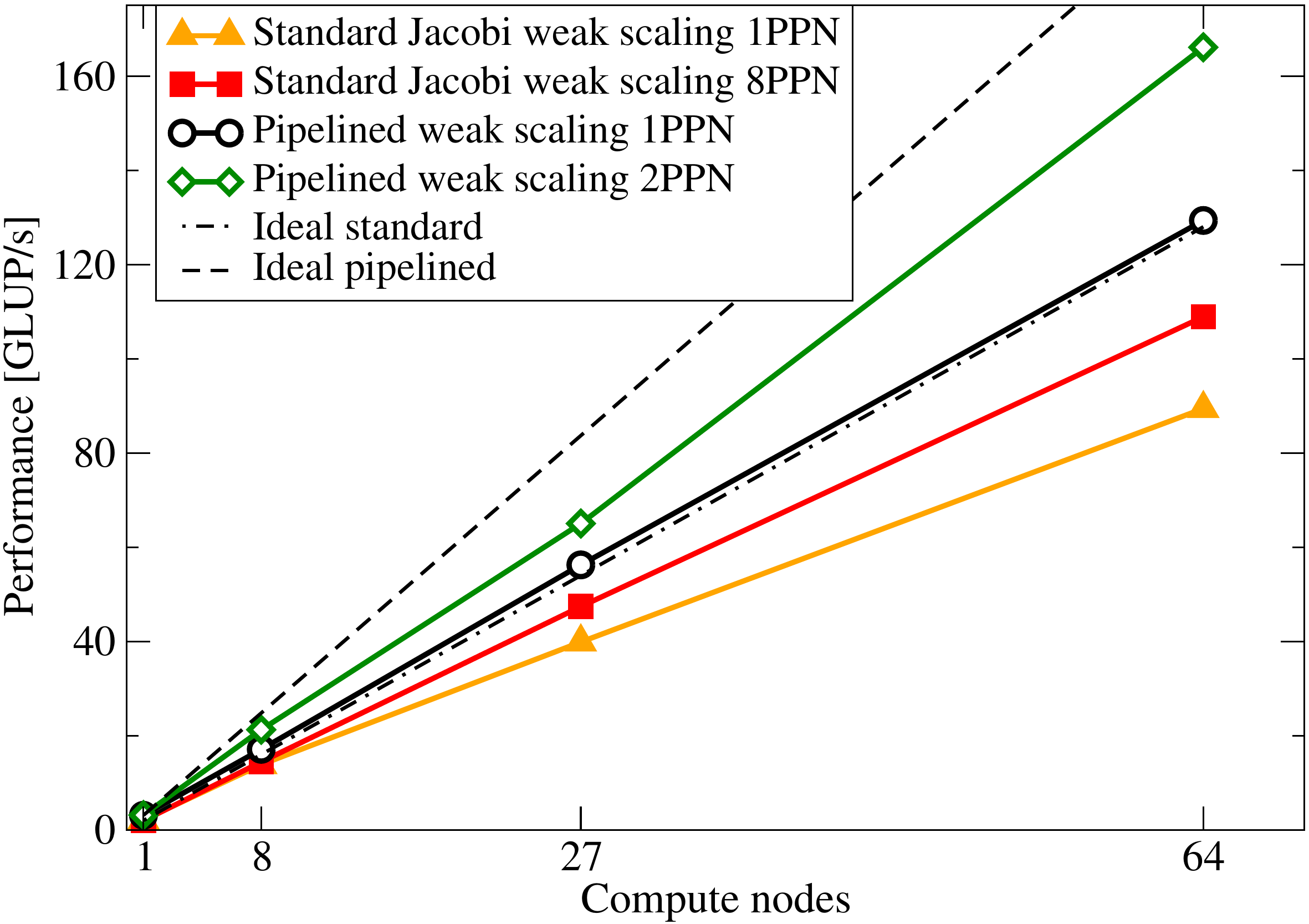}
\caption{\label{fig:dm-weak}Distributed-memory parallel performance
	(weak scaling) of the standard and the multi-halo
        pipelined Jacobi solvers with relaxed
	synchronization, at a problem size of $600^3$ per process.}
\end{figure}
Implementation of the hybrid MPI/OpenMP pipelined Jacobi code was
straightforward, with no explicit or implicit overlapping of
communication and computation. The MPI library used (Intel MPI
3.2.2.006) does not support asynchronous non-blocking transfers.
Profiling has shown that copying halo data from boundary cells to and
from intermediate message buffers causes about the same overhead as
the actual data transfer over the QDR-InfinBand interconnect.
We have investigated the performance of our distributed-memory
multi-halo pipelined Jacobi solver in strong and weak scaling
scenarios, respectively, and compared to the standard solver.

Figure~\ref{fig:dm-strong} shows performance data for strong
scaling at a constant overall problem size of $600^3$
on 1 to 64 nodes. The standard Jacobi
code (triangles and squares) was run with one MPI process per
core (8PPN) and in ``hybrid vector'' mode (1PPN),
the latter being clearly inferior. The ``2PPN'' multi-halo pipelined
version, which uses one MPI process per socket, outperforms
the ``1PPN'' version considerably because of improved NUMA
placement. However, it is just a slight progress compared
to the standard Jacobi code, and most of the performance
gain achieved by pipelined blocking on the node level is lost
due to the dominating communication overhead.
This is exactly in line with the results from the multi-halo
performance model in the previous section.

The situation changes with weak scaling. Figure~\ref{fig:dm-weak}
shows weak scaling data at a problem size of $600^3$ per node.
Again, although communication is now less important,
the hybrid vector version of the standard solver performs
worse than the pure MPI version. The multi-halo code with
one MPI process per socket, however (2PPN), can now maintain
about 80\% of the single-node speedup also in the distributed-memory
parallel case. Note that this results depends on the problem size
and the communication network; on a slow network, or with a
much smaller problem size per node, communication may still
play a dominant role and cancel out the gain from temporal
blocking.

\section{Summary and outlook}

We have demonstrated that multicore-aware pipelined temporal blocking
can lead to a substantial performance improvement for the Jacobi
algorithm on current multicore architectures (Intel Nehalem EP/EX, 
AMD Istanbul)\@.
Substitution of a global barrier by relaxed synchronization between
neighboring threads adds to the benefits. A simple bandwidth-based
performance model was shown to be insufficient to predict 
the expected speedup at low thread counts. Instead,
a cycle-by-cycle analysis of inter-cache traffic for the Jacobi
algorithm revealed that instruction execution and cache bandwidth 
rather than memory bandwidth are the limiting factors in those cases.

In comparison to earlier, more bandwidth-starved processors,
the potential gain on Nehalem EP is limited due to its high memory
bandwidth and relatively slow L3 cache.  However, future multicore
processors can be expected to be less balanced, and thus profit more
from temporal blocking, especially if the shared outer-level cache is
scalable. Our Nehalem EX test system could serve as a
``blueprint'' of such a design; its low memory bandwidth and very fast
L3 cache make temporal blocking extremely beneficial. The performance
characteristics of the AMD Istanbul processor, although it is on par
with Nehalem EP by raw numbers, are poorly understood and 
worth investigating further.

We have also shown, theoretically and in practice, under
which circumstances it is possible to port the temporal blocking
speedup to a distributed-memory parallel (hybrid) code.  A hybrid,
temporally blocked lattice Boltzmann flow solver based on the
principles presented in this work is under development.

Further optimizations are possible: One main drawback of our method is
that all cores in a node form a single large pipeline, inhibiting
optimal ccNUMA placement. This could be corrected by a domain
decomposition strategy similar to the one demonstrated in
Ref.~\cite{wellein09}, but experience shows that in practice (e.g., if
a pipelined stencil code should be used as a smoother component in
a Multigrid algorithm) it is usually best to use one MPI process
per ccNUMA domain anyway.


\section*{Acknowledgment}

Lively discussions with Darren Kerbyson, Thomas Zeiser, and Johannes
Habich are gratefully acknowledged. We are indebted to Intel
for providing early access hardware and technical information. This
work was supported by BMBF under grant No.\ 01IH08003A (project
SKALB), and by the Competence Network for Scientific High Performance
Computing in Bavaria (KONWIHR) through the project OMPI4PAPPS.

\small\RaggedRight

\end{document}